# 3D Segmentation of Neuronal Nuclei and Cell-Type Identification using Multi-channel Information

Antonio LaTorre*§[1,2], Lidia Alonso-Nanclares∗§[3,4], José María Peña[1,2] and Javier DeFelipe[3,4]

[1]DATSI, Escuela Técnica Superior de Ingenieros Informáticos, Universidad Politécnica de Madrid, Campus de Montegancedo s/n, Boadilla del Monte 28660 Madrid, Spain.

[2]Center for Computational Simulation, Universidad Politécnica de Madrid, Campus de Montegancedo s/n, Boadilla del Monte 28660 Madrid, Spain.

[3]Instituto Cajal (CSIC), Avda. Doctor Arce 37, 28002 Madrid, Spain.

[4]Laboratorio Cajal de Circuitos Corticales (CTB), Universidad Politécnica de Madrid, Campus de Montegancedo s/n, Pozuelo de Alarcón 28223 Madrid, Spain.

§Both authors contributed equally to this work

**\*Corresponding authors:**

Antonio LaTorre (email: a.latorre@upm.es)

Lidia Alonso-Nanclares (email: aidil@cajal.csic.es)

**Other authors emails:**

José-María Peña (email: jm.penya@lurtis.com)

Javier DeFelipe (email: defelipe@cajal.csic.es)

**ORCID:**

Antonio LaTorre: 0000-0002-8718-5735

Lidia Alonso-Nanclares: 0000-0003-2649-7097

Jose Maria Peña: 0000-0002-3303-0455

Javier DeFelipe: 0000-0001-5484-0660

**Highlights**

- Automatic detection of cells is critical in neuroanatomical studies
- We present a tool for automatic segmentation that allows cell type discrimination
- This tool provides data to study the number of cells and their spatial distribution

**Abstract (250max)**

<u>Background:</u> Analyzing images to accurately estimate the number of different cell types in the brain using automatic methods is a major objective in neuroscience. The automatic and selective detection and segmentation of neurons would be an important step in neuroanatomical studies.

<u>New Method</u>: We present a method to improve the 3D reconstruction of neuronal nuclei that allows their segmentation, excluding the nuclei of non-neuronal cell types.

<u>Results</u>: We have tested the algorithm on stacks of images from rat neocortex, in a complex scenario (large stacks of images, uneven staining, and three different channels to visualize different cellular markers). It was able to provide a good identification ratio of neuronal nuclei and a 3D segmentation.

<u>Comparison with Existing Methods</u>: Many automatic tools are in fact currently available, but different methods yield different cell count estimations, even in the same brain regions, due to differences in the labeling and imaging techniques, as well as in the algorithms used to detect cells. Moreover, some of the available automated software methods have provided estimations of cell numbers that have been reported to be inaccurate or inconsistent after evaluation by neuroanatomists.

<u>Conclusions</u>: It is critical to have a tool for automatic segmentation that allows discrimination between neurons, glial cells and perivascular cells. It would greatly speed up a task that is currently performed manually and would allow the cell counting to be systematic, avoiding human bias. Furthermore, the resulting 3D reconstructions of different cell types can be used to generate models of the spatial distribution of cells.

## 1. Introduction

Analyzing images to accurately estimate the number of different cell types in the brain using automatic detection methods is a major objective in neuroscience. Some manual methods — such as unbiased stereological cell counting— are reliable and effective to obtain estimations of brain cell numbers in selected brain regions of interest (for a review, see Schmitz et al., 2014). However the time-consuming nature of this type of method leads to bottlenecks and is not completely error free; even though the counting parameters have been adjusted to minimize the coefficient of error, it still exists (e.g., Schmitz and Hof, 2000). Thus, cell number estimates always vary and the range of such variation needs to be known. Automatic detection of neurons and glial cells would be an important step in a large number of neuroanatomical studies, not only to estimate the number of different cell types, but also to obtain three-dimensional (3D) reconstructions that can be used to generate models of the spatial distribution of cells.

There is a relatively large number of software tools specifically developed to perform these studies (reviewed in Meijering et al., 2012; Schmitz et al., 2014; Toyoshima et al., 2016). However, the problem is that different automatic methods yield different cell counts in the same brain regions, due to differences in the labeling and imaging techniques, as well as in the algorithms used to detect cells (e.g., Wu et al., 2000; Lin et al., 2003; Herculano-Houzel, 2005; Oberlaender et al., 2009; 2012; Toyoshima et al., 2016; Kelly and Hawken, 2017; Ruszczycki et al., 2019; Grein et al., 2020). Furthermore, some methods over- or under-estimate the number of objects automatically detected, although the error between over-and under-segmented objects is usually compensated for. Some of the available automated software methods have provided estimations of cell numbers in brain tissue that have been reported to be inaccurate or inconsistent after evaluation by neuroanatomists (Schmitz et al., 2014). It is therefore very difficult to evaluate and compare the results obtained by different laboratories.

Moreover, other authors, including ourselves, are interested in both counting and correctly identifying each object (LaTorre et al., 2013a; Kelly and Hawken, 2017). Furthermore, most of the available software tools do not provide 3D characteristics of the individually segmented cells to be further analyzed — characteristics such as centroids and Feret diameter, among others. Therefore, it is necessary to have tools which perform 3D cell segmentation to later visualize and analyze such features.

The simultaneous staining of different cell types and the improvement of immunofluorescent staining methods (e.g., Leichner and Lin, 2020) offers an excellent opportunity to analyze the cellular composition and their spatial distribution in any region of the brain. In the present work, we have developed an algorithm that can be used together with other segmentation tools to improve segmentation and data extraction to study not only the number of neurons, but also their spatial distribution. A crucial step has been added to the segmentation pipeline: detection and removal of blood vessels and the adjacent perivascular cells, which represent a major structural component of the brain tissue. This is a very important feature indeed since, despite blood vessels and their perivascular components (astrocytes and microglia; endothelial cells; and pericytes) being very abundant and widely distributed throughout the brain (Andreone et al., 2015), in most studies these perivascular components are not considered or are included as glial cells. The issue with this is that, in general, perivascular cells are indistinguishable from other non-neuronal cells and their nuclear staining interferes with the automatic detection/segmentation tools.

The present algorithm can be applied as an additional step to any detection and/or segmentation software to enhance the identification and quantification of different cell types in stained brain sections. The present study describes in detail the capabilities of this algorithm applied to a real scenario involving 3D segmentation of neuronal nuclei performed in confocal stacks of images acquired from rat neocortex slices that have been triple-labeled to visualize all cell nuclei, as well as selective visualization of neuronal nuclei and blood vessels.

## 2. Material and Methods

### 2.1. Tissue Preparation

For this study, 6 male Wistar rats were sacrificed on postnatal day 14, by administration of a lethal intraperitoneal injection of sodium pentobarbital (40 mg/kg). The animals were immediately intracardially perfused with 4% paraformaldehyde in 0.1 M phosphate buffer. The brain was then extracted from the skull, fixed overnight and sliced into coronal sections (50μm) that were collected serially. All animals were handled in accordance with the guidelines for animal research set out in the European Community Directive 2010/63/EU, and all procedures were approved by the local ethics committee of the Spanish National Research Council (CSIC).

Staining of the brain tissue must be considered as a critical step for the cell segmentation. Triple labeling for all cell nuclei (DAPI), the neuronal nuclei (anti-NeuN) and the blood vessels (anti-RECA-1) was performed. However, the set of antibodies and markers to simultaneously label these elements must be considered before the analysis, since the tissue fixation and processing might involve some technical constraints. Optimized staining methods can be applied to simultaneous fluorescent staining (Leichber and Lin, 2020). What follows is a description of the protocols used after testing several pilot experiments.

Sections containing the hindlimb region of the somatosensory cortex (S1HL; by Paxinos and Watson, 2007) were stained using immunofluorescence co-localization staining. Free-floating sections were incubated for 2 h in blocking solution: phosphate buffer (PB: 0.1M) with 0.5% Triton-X and 3% normal goat serum (Vector Laboratories, Burlingame, CA, USA). The sections were then incubated overnight at 4ºC with a mixture of rabbit anti-neuron-specific nuclear protein (NeuN, 1: 2000, Chemicon, Temecula, CA, USA) and a mouse anti-endothelial cell antibody (RECA-1, 1:2000, Abcam plc, Cambridge, UK). After rinsing in PB, the sections were incubated for 1 h at room temperature with Alexa Fluor 647 goat anti-mouse (1:1000, in blocking solution; Molecular Probes, Eugene, OR, USA) and Alexa Fluor 594 goat anti-rabbit (1:1000). Thereafter, the sections were stained with a solution containing 105 mol/L of the fluorescent dye 4, 6-diamidino-2-phenylindole (DAPI; Sigma D9542, St Louis, USA). After staining, the sections were mounted with ProLong Gold Antifade Reagent (Invitrogen, Carlsbad, CA, USA).

## 2.2. Image Acquisition

Sections were examined with a Zeiss 710 confocal laser scanning system (Carl Zeiss Microscopy GmbH, Jena, Germany). NeuN-immunoreactivity (for neurons), RECA-1-immunoreactivity (for blood vessels) and DAPI staining (for nuclei of all cell types) fluorescence was recorded through 3 separate channels. Confocal image stacks of 140-150 z-planes were obtained with an EC PL NEO 40x immersion lens (N.A. 1.3), using a z-step of 0.21μm and a scanning resolution of 1024x1024 pixels (pixel size 0.21μm). Acquisition of final images containing all cortical layers was performed using the "Tile Scan" tool, with an overlap of 10% and a final stitching processing step, using the software ZEN (2012, SP1 v8.1, Carl Zeiss Microscopy). Final images comprise about 16-18 tiles, producing a final stitched file of about 3GB (including the 3 channels and all the tiles). Throughout this manuscript, the term 'cortical column' refers to the collected final images, from cortical surface to grey/white interface. An example of an acquired confocal image stack of the somatosensory cortex in a triple labeled brain section is shown in Fig. 1.

## 2.3. Description of the code

Current implementation was done in Matlab and needed the Image Processing Toolbox (licensed by MatLab®). Additionally, it uses the "export_fig" and the "slicer" Matlab packages, provided in the "libs" folder, which are freely distributed according to their licenses. Finally, in order for the algorithm to work, you need to download and copy the MIJ library ("mij.jar" and "ij.jar" files) from [http://bigwww.epfl.ch/sage/soft/mij/](http://bigwww.epfl.ch/sage/soft/mij/) and copy them to the "libs" folder (Sage et al., 2012).

The main entry point for the "3D Cells Segmenter" is the "demoCS3DGUI.m" file ([http://cajalbbp.es/storage/3D_Cell_Segmenter_Source_Code.zip](http://cajalbbp.es/storage/3D_Cell_Segmenter_Source_Code.zip)). This program presents a GUI that can be used to select the stacks that are going to be processed for each of the channels, the output directory for the resulting files with the segmentation and some of the basic parameters of the algorithm. All the remainder parameters can be configured from the "Config.m" file according to the guidelines provided in Table 1.

The results reported in this work have been obtained with a conventional computer (Intel Core i7m 3.4Ghz CPU 16GB RAM) using Ubuntu Linux as operating system and MatLab R2011b as programming language. All the parameters have been chosen in an experimentally conservative way (as summarized in Table 1).

For example, the parameters used in the division of clusters of cells by analyzing the centroids have been selected so that cells which have already been properly segmented are not over-segmented by the use of this improvement.

**Table 1.** Parameter values used

| | |
|---|---|
| Similar overlapping threshold $\delta_1$ | 0.2 |
| Minimum overlapping threshold $\delta_2$ | 0.2 |
| k-means iterations *maxClusters* | 3 |
| Minimum size of cluster $\alpha$ | 3 |
| Size ratio for clusters $\beta$ | 0.3 |
| Maximum overlapping of clusters $\gamma$ | 0.65 |

*2.4. Manual Validation*

Manual segmentation was performed by experts in neuroanatomy in the same files that were processed using our segmentation algorithm to validate the accuracy of the results. For this purpose, manual segmentation of the labeled neuronal nuclei (NeuN-ir) and cell nuclei (DAPI) was performed in all cortical layers (I, II, II, IV, V and VI) using EspINA software (EspINA Interactive Neuron Analyzer, 2.1.9; https://cajalbbp.es/espina/), which allows the segmentation of objects in a reconstructed 3D volume (for a detailed description, see Morales et al., 2011). Manually segmented cell nuclei were individually tagged and unequivocally identified using EspINA, allowing further validation of the automatic segmentation obtained by our algorithm.

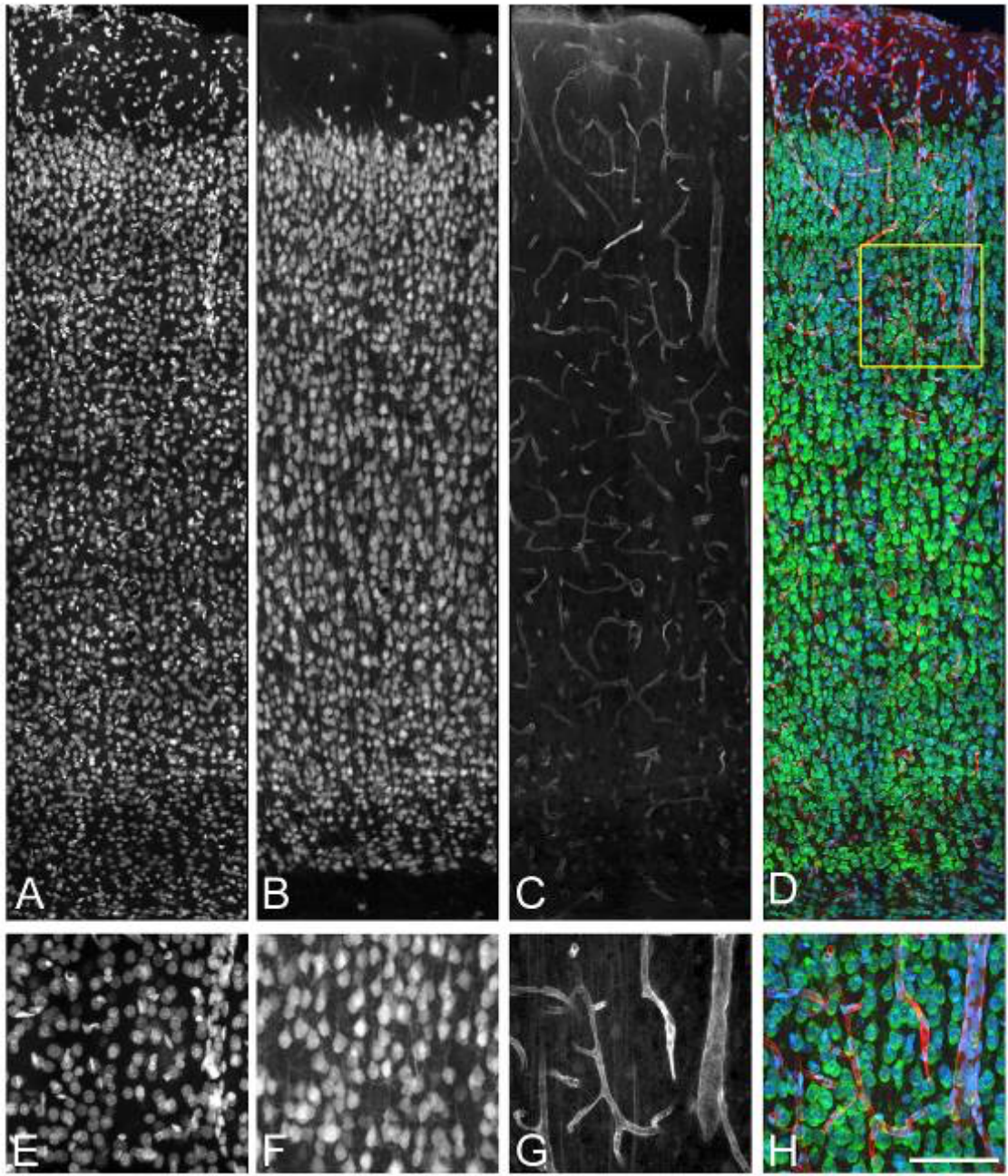

**Figure 1.** Confocal stack z-projection images (29 μm-thick) taken from the same field of a cortical column of the S1HL rat cerebral cortex in a triple-labeled section. **A:** DAPI staining of all cell nuclei; **B:** NeuN-immunoreactivity for neurons; **C:** RECA-1-immunoreactivity for blood vessels. **D:** Overlay of the 3 channels allows simultaneous visualization of all markers: cell nuclei (blue), neuronal soma (green) and blood vessels (red). **E-H** Higher magnification of the boxed region indicated in **D,** respectively. In **H**, numerous cell nuclei can be observed adjacent to blood vessels. Scale bar (in **H**): 150μm (**A-D**), and 75 μm (**E-H**).

*2.5. Digital Illustrations*

Figures were prepared digitally using Photoshop CS4 software (Adobe Inc., San Jose, CA, USA). Confocal images were acquired using a Zeiss 710 confocal laser scanning system (Zeiss). Z-stack projection images were produced using FIJI software (Image J 1.51h, NIH, USA). Snapshots of the 3D visualizations and representation of the segmented nuclei were performed with EspINA software. The contrast, brightness and sharpness of images were adjusted as needed for each plate.

## 3. Results

In this section, we review some of the problems that were traditionally encountered and outline how they have been addressed in this new 3D segmentation algorithm (see software description section).

Briefly, in our previous study (LaTorre et al., 2013a), we proposed a 3D segmentation algorithm that was able to reconstruct 3D volumes of cells from 2D segmentations that could be computed with any other algorithm. Under-segmentation and over-segmentation issues were corrected by a number of additional post-processing filters. The result of that 3D reconstruction algorithm was a set of labeled 3D cell nuclei. Our algorithm was very efficient at obtaining results in medium to small problems (in terms of size files; confocal images stacks of less than 500MB were used). However, to obtain an accurate segmentation of whole extent of cortical columns, using confocal image stacks for 3 simultaneous channels, a number of issues have been overcome, both from a computational and an algorithmic point of view.

As a first positive result, the new algorithm does process large files, such as those confocal stacks containing the entire extent of gray matter from a given cortical column. Changes are highlighted in Fig. 2, which contain the workflow of the main steps of the new algorithm. Secondly, the inclusion of larger scanned brain regions and an additional channel (blood vessel staining) provides better discrimination of the cellular types whilst at the same time allowing the study of a cortical column in its entirety. Finally, validation results displayed similar values of correctly labeled neuronal nuclei as our previous version.

As a proof of concept, confocal stacks of images were used to produce a single composite stack of a whole cortical column (from layer I to layer VI). The image stacks were scanned with the confocal microscope via 3 channels (as described in section 2.2.), producing files of 3–4.5 GB. Since the stacks comprise all cortical layers, they contain layers with diverse density, distribution and sizes of neurons, as well as a large number of labeled vascular and perivascular cells. Thus, the stacks can be considered as representative samples to apply and test the algorithms (Fig. 1).

*3.1. Computational improvements*

As mentioned above, some of the main problems were related to the challenges involved in scaling up the algorithm. In particular, we had to control the memory footprint of the algorithm and improve the efficiency of the algorithm in terms of runtime.

To deal with the first issue, the current code minimizes the number of (partial) copies that we store simultaneously in the memory. For the computational efficiency, we have replaced loop structures by vectorized implementation whenever possible, which implied, in some cases, that we had to rewrite some important parts of the code. This has provided an improvement of several orders of magnitude in the execution time of some parts of the algorithm code and, thus, it allowed the processing of some stacks of images that simply took too long for the previous version of the algorithm to finish. In addition, these computational improvements have shown that our algorithm was able to troubleshoot freezing issues in large datasets which were impossible to solve using other software tools (see section 3.7).

*3.2. Addition of the blood vessels channel*

The inclusion of a third channel containing the labeling for blood vessels allows us to better discriminate among different types of cells. This information facilitates not only the separation of neurons from non-neuronal cells in general, but also the identification of perivascular cells. There are several histological methods to label the blood vessels in mammalian brain. In our case, an optimal staining of perfused rat brain slices was achieved by immunostaining for an anti-endothelial cell antibody (RECA-1; as described in Methods), which results in a general labeling of the vascular network in brain sections, and permits visualization of blood vessels (Fig. 1C).

An additional step of removing any type of perivascular cells located in close apposition to the blood vessels selectively discarded numerous non-neuron cells, thereby improving the segmentation of neurons via a reduction in the number of objects to be considered (Figs. 2, 3).

*3.3. Improvements aimed at reducing over-segmentation*

One of the problems identified in our previous study was that an inaccurate 2D binarization normally leads to over-segmentation when the 3D algorithm is applied. In the present study, with large stacks of images, new situations arouse requiring specific solutions.

*3.3.1. Removal of small objects:*

After trying to merge small objects with other cells, some other small objects remain in the images, especially when the stack grows in size and the staining of the tissue is not homogeneous in all axes (commonly observed in the z-axe, due to limited antibody penetration). In these cases, we apply a series of filters to remove those objects that are incomplete or fragments of labeled cells, which are considered as artifacts (Fig. 2 Panel 2). In most of the cases, these small objects are just noise coming from the uneven illumination of the image, or auto-fluorescence artifacts. In other cases, they can correspond to the first or the last slices of a cell that, due to a discontinuity in the original image, could not be assigned to the soma of the cell. This can have a marginal impact on the reported cell size of the main object but it has a profound impact on the overall estimation of the number of cells. In our experiments we have imposed a minimum number of 3 planes (in the Z axis) to consider the object as a cell nuclei, and a minimum volume of 10,000 voxels. This corresponds to an object with an average volume of approximately 93 $\mu m^3$, matching an average radius of 4.5 $\mu m$, which is in line with previously reported neuronal body sizes in rodents (Henderson et al., 1980; Gittins and Harrison, 2004), and with the object size that we are considering.

*3.3.2. Improvement of the second pass of the binarization process:*

In LaTorre et al. (2013b), we proposed a 2D image binarization method consisting of two steps: a global and a local step. The former involved a global threshold that is applied to the whole image to try to identify all the existing objects in the image, whereas the latter processes each of the individual objects detected in the previous step to refine the contour of the cells.

In this new version of the algorithm, we have improved this 2-step binarization method by tackling several problematic situations that may happen when using the original version.

An opening operation (highlighted step in Fig.2 panel 2) is applied at the end of the second pass of the binarization to remove noisy pixels that might have appeared after this second pass. This can happen in some areas of the stack of images when trying to soften the contour of a cell. However, this second pass of the binarization process might introduce some foreign objects coming from other cells within the "region of interest" (ROI) that we are processing to soften the contour of the cells. In these cases, one or more extra (smaller) objects would appear in the ROI that may or may not overlap completely with the binarization of the other cell. If this overlapping is not complete, some small objects will appear as independent cells, increasing the over-segmentation ratio. For this reason, the second pass of the binarization method allows only one object (the main one) to be returned. All the other objects that might have appeared in the ROI are removed (highlighted step in Fig.2, panel 2).

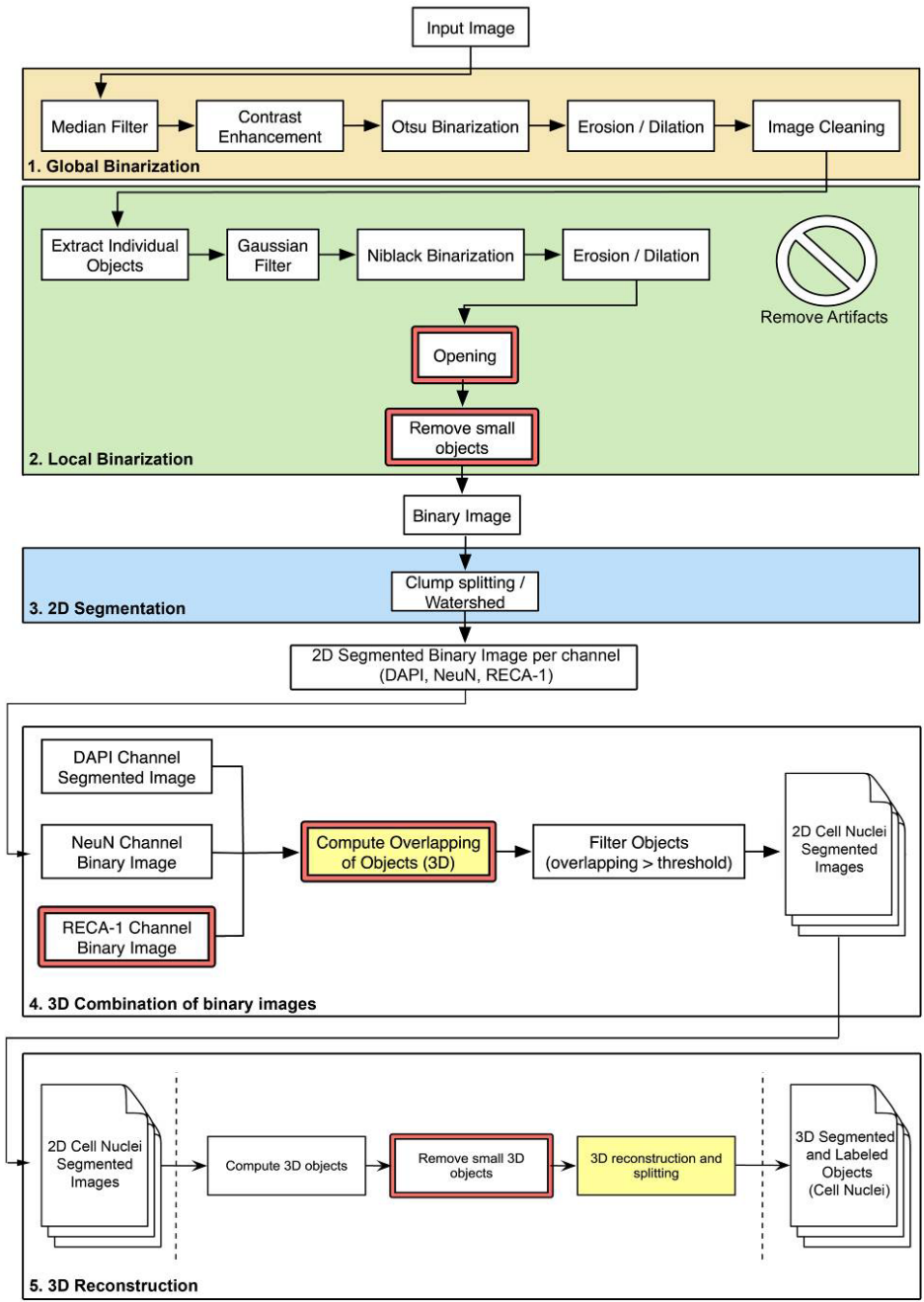

**Fig. 2** General overview of the pipeline to illustrate the steps from the input (3-channels confocal image stacks) to the output (3D segmented and labeled objects):

**(1)** Global Binarization step. **(2)** Local Binarization: "Opening" and "Removal of small objects" filters at the end of the Local Binarization process have been added to the current workflow to improve the final 2D segmented binary images. **(3)** 2D Segmentation: produces the 2D segmented binary images. (**4**) 3D Combination of binary images: 3D combination of the 2D segmented binary images using the three channels, computing an "Overlapping of Objects in 3D", which produces a final set of 2D cell nuclei segmented images. Each set of images contains the cell nuclei of different cell populations identified from the additional channels used at this step. (**5**) 3D Reconstruction: after the "Compute 3D object" step, we added the "Remove small 3D objects" as a new step. This step produces the final 3D labeled objects (cells nuclei; different cell nuclei populations to be segmented can be chosen depending on the channels used in the previous step). Moreover, enhancements in vectorization and other computational improvements have been added to the final step of "3D reconstruction and splitting". New steps are highlighted with red frame and steps highlighted in yellow have been greatly improved in terms of computational effort and memory usage.

### *3.3.3. Avoid splitting of 3D blobs in slices with small objects:*

As described in LaTorre et al. (2013a), the 3D segmentation algorithm divided a 3D blob into multiple blocks whenever it found that, in some slices, there were multiple cells identified in 2D. However, this can be problematic if the object sizes are very dissimilar: this frequently means that one or more of the objects actually do not represent a new cell but a fragment of the main cell that could not be removed by the previous filters. In these situations, if the size of all the pieces is not over a prefixed threshold, we avoid splitting the blob at that slice and continue processing the next one (Fig. 2, panel 3).

### *3.3.4. 3D Merging of channels:*

In the present experiments, staining was not homogeneous along the full extent of the cortical column, and we have also included an extra channel (for blood vessels labeling; Fig. 1). Thus, we have incorporated a new 3D combination approach (Fig. 2, panel 4, highlighted). This method works in a similar way to its 2D counterpart but, instead of computing 2D overlapping (in pixels), we compute 3D overlapping (in voxels; highlighted in Fig. 2, panel 4). Moreover, this step of the algorithm allows cell populations to be identified as different cell types according its labeling.

### *3.4. Improvements designed to reduce under-segmentation*

To avoid under-segmentation, we correct some extreme cases in which the second pass of the binarization process could make the cell grow towards the borders of the sub-region where the cell is located. This is a rare case and only happens when one DAPI-labeled object (a cell nucleus) is tightly coupled with another shiny object. However, we take these cases into account and abort the second pass if we detect that this is happening, keeping the original binarization (i.e., the one from the first, global pass).

*3.5. Other improvements*

In this new version of the algorithm, we no longer needed a specific filter to remove shiny objects tightly coupled to other cells, since the inclusion of the blood vessels channel helps to discard all such cases (Figs. 3, 4).

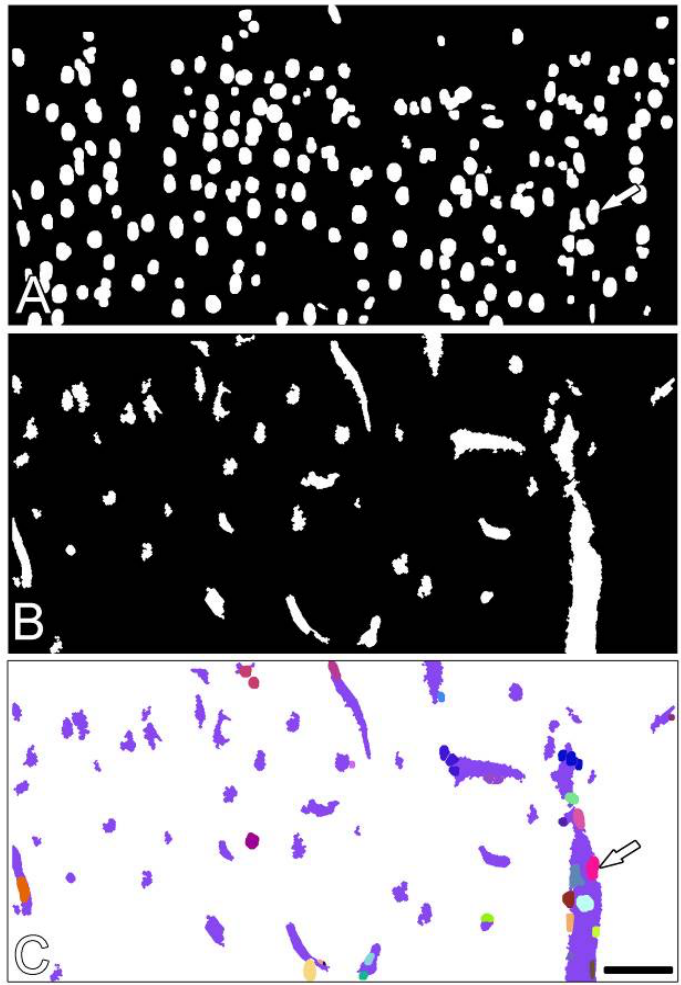

**Fig. 3 a, b** Binary images (from a single plane, z=0.21 μm) computed from the DAPI channel (**a**), blood vessel channel (**b**) **c** Combination of 2D segmented non-neuronal cell nuclei and blood vessels channels allows the removal of non-neuronal cells to refine segmentation of neuronal nuclei. Arrows (in **a** and **c**) indicate a non-neuronal cell nucleus adjacent to a blood vessel which will be removed by the algorithm. Note the large number of perivascular cell nuclei. Scale bar (in **c**): 40μm.

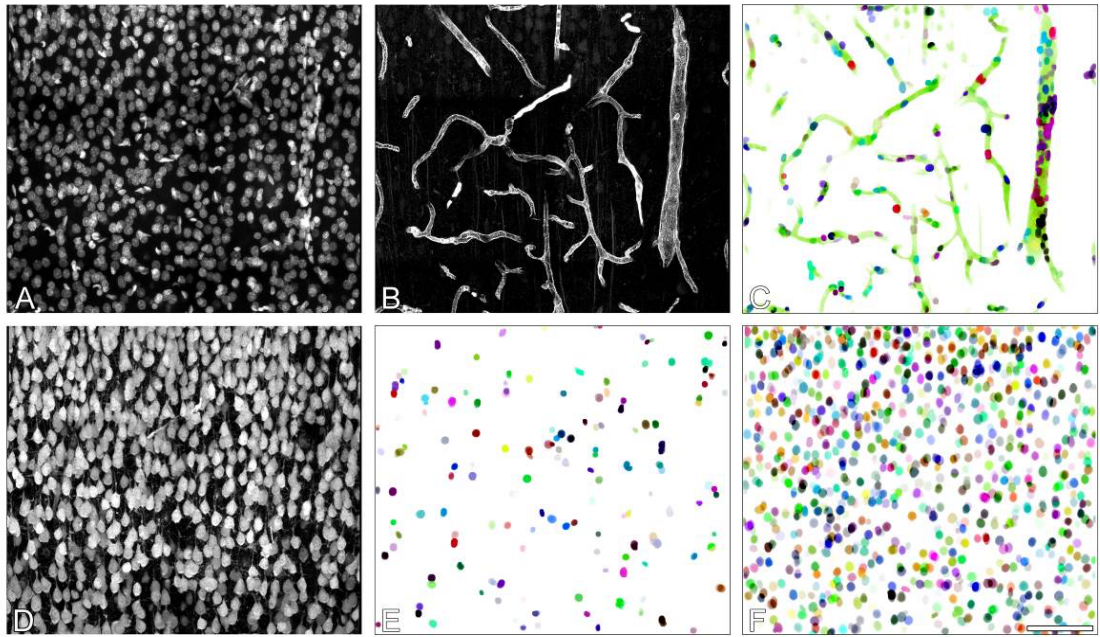

**Fig. 4** Plate to illustrate 3D channel combination and segmentation performed by the algorithm in cortical layers II - III of the S1HL rat neocortex. **a, b** Z-projection images of confocal stacks with DAPI-stained nuclei for all cell types (**a**) and RECA-1- immunoreactivity for blood vessels (**b**). **c** Binary image obtained by combining **a** and **b** to illustrate nuclei of vascular and adjacent cells as well as a segmented blood vessel. These segmented cell nuclei (**c**) are removed in the next step to refine segmentation. **d** Z-projection image of a confocal stack showing NeuN-immunoreactive neurons. **e** Segmented nuclei of non-neuronal cells (not stained for NeuN) obtained by combining **c** and **d**. **f** Final segmentation of neuronal nuclei obtained by the algorithm, after removal of blood vessels, perivascular cells (**c**) and other non-neuronal cells (**e**). Thus, only neurons are segmented. Scale bar (in **f**): 75 µm

### 3.6. Validation results

The accuracy of neuronal nuclei detection in stained histological brain sections varies depending on diverse factors such as the characteristics of the nuclei (shape and size), as well as the visualization method (e.g., immunofluorescence) and image acquisition technique used (Toyoshima et al., 2016).

Manual validation of the accuracy of the revised algorithm was performed in all cortical layers (I, II, II, IV, V and VI). 12 image stacks (140–150 z-planes, corresponding to a z-thickness of around 30 µm) from 3-channel laser confocal microscopy (stained for blood vessels, cellular nuclei and neuronal nuclei), were used for the manual validations. A total of 3,589 segmented neuronal nuclei were manually validated (56 from layer I, 480 from layer II, 806 from layer III, 420 from layer V, 733 from layer V and 1094 from layer VI). The resulting validated dataset is summarized in Table 2.

In summary, the percentage of correct automatic segmentations achieved by the current version of our algorithm was around 93%, reaching a maximum value of approximately 97% in layer I and a minimum of 89% in layer V.

**Table 2. Manual validation results.** For each labeled 3D neuronal nucleus, the expert determines if it corresponds to a properly segmented neuronal nucleus or if it matches one of the four errors considered. Validations were performed in different cortical layers.

|  | Cortical Layer | | | | | |
|---|---|---|---|---|---|---|
|  | I | II | III | IV | V | VI |
| **Correct** | 96.77% | 94.02% | 94.37% | 90.00% | 89.24% | 92.19% |
| **Type-1 error: Over-segmentation** | 3.23% | 2.33% | 2.89% | 6.07% | 6.22% | 3.24% |
| **Type-2 error: Under-segmentation** | 0% | 2.99% | 2.09% | 3.57% | 2.02% | 1.79% |
| **Type-3 error: Noise detected as cell** | 0% | 0.66% | 0.64% | 0.36% | 2.52% | 2.79% |
| **Type-4 error: Undetected cell** | 0% | 0% | 0% | 0% | 0% | 0% |

3.7. Comparison with other segmentation methods

Finally, we compared the performance of our current algorithm with seven previously published segmentation tools: "CellSegmentation3D", which uses gradient flow tracking techniques and was developed for clump splitting (Li et al., 2007); Ilastik, a tool based on machine learning techniques and which uses image features (Sommer et al., 2011); FARSight, based on graph cut techniques (Al-kofani et al., 2010); the 3D watershed plugin in ImageJ, consisting of local peak detection and seeded watershed (Ollion et al., 2013) applied after our binarization step; the method published by Toyoshima et al. (2016), which is based on the curvatures of the iso-intensity surfaces. Recent tools were also included: StarDist, based on star-convex polygons for object detection (Schmidt et al., 2018;Weigert et al., 2020) and Vaa3D, a visualization-assisted analysis system (Xu and Prince, 2014).

When we tried to apply these segmentation tools to our original confocal stacks of images containing the whole cortical column, the tools crashed during processing or took too long to compute. As described above, our present version of the algorithm contains computational improvements that avoid these crashing issues that occur with large stacks of images. Thus, datasets used for validation were cropped stacks obtained from our original images stacks previously used for the validation. Cropped stacks were limited to a width of 50 μm and a height of 50 μm (maintaining the original resolution), and the entire thickness in the z-axis.

Regarding the initial result, the CellSegmentation3D tool was unable to process the cropped stacks, so it was discarded from comparisons. In general, the compared methods had difficulty handling large 3D images with densely packed objects (Table 3). Only StarDist and Vaa3D gave cell numbers close to those provided by our algorithm. However, these tools resulted in sub-segmentation and over-segmentation errors. In addition, none of the tools were designed for cell-type discrimination; they produced a single result which included all cell types, whereas our algorithm computed a refined segmentation, with the capacity to distinguish neurons between the cell types (as indicated in Table 3). This deficiency regarding discriminative segmentation is due to the lack of the simultaneous use of channels providing different information about the segmented cells. In our current version of the algorithm, the main value comes from refinement by discarding the labeled objects which were also present in the blood vessel (RECA1) channel and the cells which were not labeled with the neuronal marker (NeuN) channel.

The results indicate that our current algorithm detects densely distributed cell nuclei in 3D with a very low false negative rate, which is the most significant improvement compared to the other methods. Moreover, the advantage of selective segmentation of neuronal nuclei improves the efficiency and accuracy of the final image analysis.

**Table 3. Comparison of segmentation results.** Number of objects (cells) automatically detected by different segmentation tools. Note that most of the tools generated a high number of over-segmentation and/or false-positive objects. StarDist misses many of the cells identified by our algorithm. Vaa3D had problems with both over- and under-segmentation, and it also missed some cells. Our current version of the algorithm also provides the number of segmented neuronal nuclei (indicated in parentheses). "-" indicates that the applied tool was unable to process the stack of images.

| Cortical layer | Current algorithm (no. of neurons) | Ilastik | FARSight | 3D watershed plug-in (ImageJ) | Toyoshima's | StarDist | Vaa3D |
|---|---|---|---|---|---|---|---|
| I | 19 (1) | 212 | 90 | 62 | 257 | 0 | 12 |
| II | 47 (37) | 903 | 982 | 174 | 862 | 21 | 38 |
| III | 37 (17) | 585 | 111 | 98 | 565 | 10 | 24 |
| IV | 21 (12) | 446 | 44 | 70 | - | 7 | 29 |
| V | 28 (15) | 928 | 25 | 110 | - | 10 | 19 |

| **VI** | 20 (16) | 326 | 40 | 82 | - | 11 | 33 |

## 4. Discussion

In recent years, a number of software tools to segment cells have been developed in the field of neuroscience (reviewed in Schmitz et al., 2014; Toyoshima et al., 2016). However, our goal involves the following additional challenges:

a) We considered the simultaneous use of 3 channels as a necessary step to selectively remove the perivascular cell subpopulation and the non-neuronal nuclei;

b) We needed to not only annotate neuronal nuclei, but also to compute the 3D information of all the segmented objects (such as the centroids, Feret diameter, etc.) to further analyze this relevant 3D information in neuroanatomical studies;

c) From our perspective, segmentations must be visualized through a GUI to correct or edit if needed.

In our case, we used EspINA software because it also allows edition and correction of the segmented objects in 3D (Fig. 5). To be more specific, we have used EspINA software because it not only has editing tools, but also has an unbiased counting frame probe which allows the estimation of the number of objects inside the counting frame volume (Morales et al., 2011). This unbiased counting frame is a regular rectangular prism bounded by three acceptance planes and three exclusion planes (Howard and Reed, 2005). We considered it important to be able to individually evaluate the segmentations made by our algorithm in order to detect or correct possible mistakes, as well as to extract data. For this purpose, our algorithm produces data in a format that is compatible for use in several software tools for the 3D visualization, edition and analysis of segmented objects.

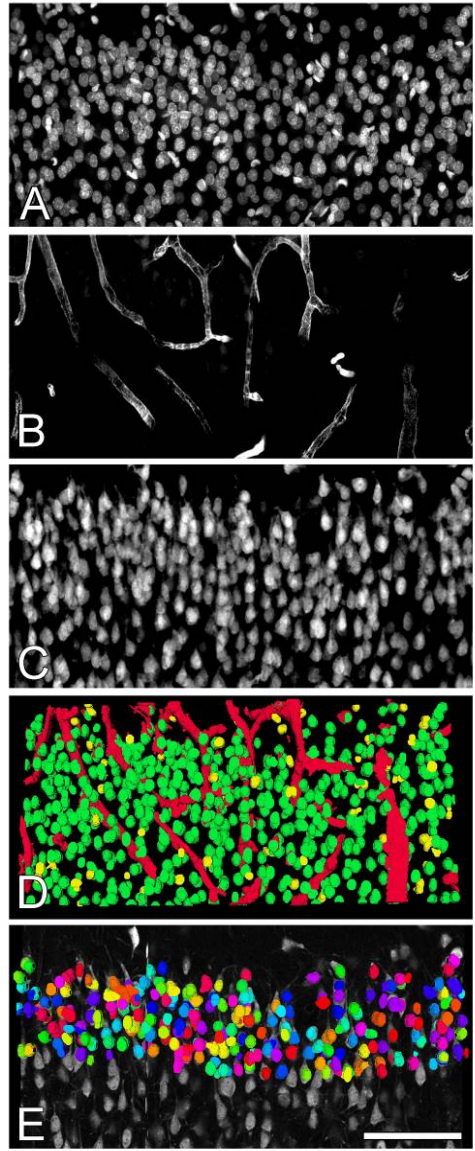

**Fig. 5** Images to illustrate the general algorithm workflow and final 3D reconstruction of neuronal nuclei in layer II of the S1HL rat neocortex as an example. **a-c** Confocal stack z-projection images showing DAPI-stained cell nuclei (**a**), RECA-1-immunoreactive blood vessels (**b**), and NeuN-immunoreactive neurons (**c**). **d** 3D Visualization of the segmented cell nuclei of neurons (in green); blood vessels and vascular cells (in red); and segmented nuclei of non-neuronal/non-vascular cells (in yellow). **e** 3D Visualization of the final segmented neuronal nuclei. Neuronal nuclei are randomly colored differently to facilitate visualization of neighboring neurons. Scale bar (in **e**): 75 μm in **a-c**, and 90 μm in **d** and **e**

As described in section 3.7, we considered some of the common methods available that could be applied to our samples. Another developed method for obtaining cell segmentation from histological brain samples is the automated algorithm proposed by Kelly and Hawken (2017), however, the tool is not freely distributed and the technical steps described in their work did not provide enough detailed information to be applied in our files.

In general, using some of the mentioned tools, computational issues regarding the file size occurred with our image stacks. The characteristics of our files did not allow the proper functioning of the tools, so the comparison was performed with smaller files, as previously described. Moreover, some of the compared tools, such as Ilastik, depend on the availability of previously segmented images containing objects to be analyzed, and only performs "density counting", that is, they count objects but do not segment them.

Although the above mentioned available tools have reported accurate results for the detection and counting of cells in the brain, they could also be improved using our algorithm as an additional step to enhance the detection of non-neuronal cells by the removal of the blood vessels and associated cells (such as pericytes). Distinguishing neuronal nuclei by refined removal of other types of cell nuclei gives our present method the advantage of higher accuracy compared to the other tools.

Finally, in general, most algorithms for automatic cell detection and counting use one (the majority) or two channels to identify labeled objects. A major contribution of the present algorithm is that it is now possible to take into account an additional image stack with an additional channel containing specific staining for blood vessels, allowing different cell populations (neurons, glia and perivascular cells) to be distinguished much more clearly. These cell populations are identified in the step 4th of the algorithm (as shown in Figs. 2, 3, 4), and they are separately processed in step 5th.

We have tested the algorithm on several stacks of images of different sizes, the largest being over 2 GB per channel — a volume of information that could not be processed by our previous version (LaTorre et al., 2013) or indeed by most of the available software tools. Even in this complex scenario (large stacks of images, uneven staining, three different channels), our algorithm provides a good identification ratio of neuronal nuclei, as validated. The present algorithm can be applied as an additional step to other algorithms that are already available, to improve the detection of neuronal cells by the removal of the numerous perivascular cells present in all brain regions.

**References**

Al-Kofahi Y, Lassoued W, Lee W, Roysam B (2010) Improved automatic detection and segmentation of cell nuclei in histopathology images. IEEE Trans Biomed Eng 57: 841–852. doi: 10.1109/TBME.2009.2035102

Andreone BJ, Lacoste B, Gu C (2015) Neuronal and Vascular Interactions. Annu Rev Neurosci 38:25-46. doi:10.1146/annurev-neuro-071714-033835.

Gittins R, Harrison PJ (2004) Neuronal density, size and shape in the human anterior cingulate cortex: a comparison of Nissl and NeuN staining. Brain Res Bull 63:155-160. doi:10.1016/j.brainresbull.2004.02.005

Grein S, Qi G, Queisser G (2020) Density Visualization Pipeline: A Tool for Cellular and Network Density Visualization and Analysis. Front Comput Neurosci 14:42. doi:10.3389/fncom.2020.00042

Henderson G, Tomlinson BE, Gibson PH (1980) Cell counts in human cerebral cortex in normal adults throughout life using an image analysing computer. J Neurol Sci 46:113-136

Herculano-Houzel S (2005) Isotropic Fractionator: A Simple, Rapid Method for the Quantification of Total Cell and Neuron Numbers in the Brain. J Neurosci 25: 2518-2521. doi:10.1523/JNEUROSCI.4526-04.2005

Kelly JG, Hawken MJ (2017) Quantification of neuronal density across cortical depth using automated 3D analysis of confocal image stacks. Brain Struct Funct 222:3333-3353. doi:10.1007/s00429-017-1382-6

LaTorre A, Alonso-Nanclares L, Muelas S, Peña JM, DeFelipe J (2013) 3D segmentations of neuronal nuclei from confocal microscope image stacks. Front Neuroanat, 7. doi:10.3389/fnana.2013.00049

LaTorre A, Alonso-Nanclares L, Muelas S, Peña JM, DeFelipe J (2013) Segmentation of neuronal nuclei based on clump splitting and a two-step binarization of images. Expert Syst Appl 40:6521-6530. doi:10.1016/j.eswa.2013.06.010

Leichner J, Lin WC (2020) Advances in imaging and analysis of 4 fluorescent components through the rat cortical column. J Neurosci Methods 341: 108792. doi.org/10.1016/j.jneumeth.2020.108792

Lin G, Adiga U, Olson K, Guzowski JF, Barnes CA, Roysam B (2003) A hybrid 3D watershed algorithm incorporating gradient cues and object models for automatic segmentation of nuclei in confocal image stacks. Cytometry A 56:23-36. doi:10.1002/cyto.a.10079

Meijering E, Dzyubachyk O, Smal I (2012) Methods for cell and particle tracking. Methods Enzymol 504:183-200. doi:10.1016/B978-0-12-391857-4.00009-4

Morales J, Alonso-Nanclares L, Rodríguez J, DeFelipe J, Rodríguez A, Merchán-Pérez A (2011) ESPINA: A tool for the automated segmentation and counting of synapses in large stacks of electron microscopy images. Front Neuroanat, 5. doi:10.3389/fnana.2011.00018

Oberlaender M, de Kock CPJ, Bruno RM, Ramirez A, Meyer HS, Dercksen VJ, Helmstaedter M, Sakmann B (2012) Cell Type-Specific Three-Dimensional Structure of Thalamocortical Circuits in a Column of Rat Vibrissal Cortex. Cereb Cortex 22:2375-2391. doi:10.1093/cercor/bhr317

Oberlaender M, Dercksen VJ, Egger R, Gensel M, Sakmann B, Hege HC (2009) Automated three-dimensional detection and counting of neuron somata. J Neurosci Methods 180:147-160. doi:10.1016/j.jneumeth.2009.03.008

Ollion J, Cochennec J, Loll F, Escudé C, Boudier T (2013) TANGO: A generic tool for high-throughput 3D image analysis for studying nuclear organization. Bioinformatics 29: 1840–1841. doi: 10.1093/bioinformatics/btt276

Paxinos G, Watson C (2007) The rat brain in stereotaxic coordinates. 6th ed. Academic Press/Elsevier, Amsterdam / Boston

Ruszczycki B, Pels KK, Walczak A, et al. (2019) Three-Dimensional Segmentation and Reconstruction of Neuronal Nuclei in Confocal Microscopic Images. Front Neuroanat. 13:81. doi:10.3389/fnana.2019.00081

Sage D, Prodanov D, Tinevez J-Y, Schindelin J. (2012) MIJ: Making Interoperability Between ImageJ and Matlab Possible. In: ImageJ User & Developer Conference, Luxembourg.

Schmitz C, Eastwood BS, Tappan SJ, Glaser JR, Peterson DA, Hof PR (2014) Current automated 3D cell detection methods are not a suitable replacement for manual stereologic cell counting. Front Neuroanat, 8. doi:10.3389/fnana.2014.00027

Schmidt U, Weigert M, Broaddus C, Myers G (2018) Cell detection with star-convex polygons. 30 In: Frangi A., Schnabel J., Davatzikos C., Alberola-López C., Fichtinger G. (eds) Medical Image Computing and Computer Assisted Intervention - MICCAI 2018 - 21st International Conference, 16-20, 2018, Proceedings, Part II, pp. 265–273. Lecture Notes in Computer Science, vol 11071. Springer, Cham. doi:10.1007/978-3-030-00934-2_30

Schmitz C, Hof PR (2000) Recommendations for straightforward and rigorous methods of counting neurons based on a computer simulation approach. J Chem Neuroanat 20:93-114

Sommer C, Straehle C, Köthe U, Hamprecht FA (2011) Ilastik: Interactive learning and segmentation toolkit. In: 2011 IEEE International Symposium on Biomedical Imaging: From Nano to Macro, 2011. pp 230-233. doi:10.1109/ISBI.2011.5872394

Toyoshima Y, Tokunaga T, Hirose O, Kanamori M, Teramoto T, Jang MS, Kuge S, Ishihara T, Yoshida R, Iino Y (2016) Accurate Automatic Detection of Densely Distributed Cell Nuclei in 3D Space. PLoS Comput Biol 12:e1004970. doi:10.1371/journal.pcbi.1004970

Weigert M, Schmidt U, Haase R, Sugawara K, Myers G (2020). Star-convex polyhedra for 3D object detection and segmentation in microscopy. The IEEE Winter Conference on Applications of Computer Vision (WACV). doi:10.1109/WACV45572.2020.9093435

Wu HS, Barba J, Gil J (2000) Iterative thresholding for segmentation of cells from noisy images. J Microsc 197:296-304

Xu C, Prince JL (2014) Gradient vector flow. In: Ikeuchi K. (eds) Computer Vision. Springer, Boston, MA. doi:10.1007/978-0-387-31439-6_712. 2014

**Compliance with Ethical Standards**

Conflict of Interest: The authors declare that they have no conflict of interest.

Ethical approval: All procedures performed in studies involving animals were in accordance with the ethical standards of the European Community Directive 2010/63/EU and local ethics committee of the Spanish National Research Council (CSIC) at which the studies were conducted.

**CRediT authorship contribution statement**

Alonso-Nanclares: Methodology, Validation, Formal analysis, Investigation, Data curation, Writing - original draft, Visualization. LaTorre: Methodology, Software, Validation, Investigation, Writing. Peña: Conceptualization, Data curation, Writing – review & editing, Supervision. DeFelipe J: Conceptualization, Data curation, Writing – review & editing, Supervision, Project administration, Funding acquisition.

**Funding**

This study was funded by grants from the following entities: Spanish "Ministerio de Ciencia e Innovación" grants PGC2018-094307-B-I00 and TIN2017-83132-C2-2-R; the Cajal Blue Brain Project (the Spanish partner of the Blue Brain Project initiative from EPFL, Switzerland); the European Union's Horizon 2020 Framework Programme for Research and Innovation under grant agreement No. 945539 (Human Brain Project SGA3).

**Acknowledgments**

The authors thankfully acknowledge the computer resources, technical expertise and assistance provided by the Centro de Supercomputación y Visualización de Madrid (CeSViMa) and the Spanish Supercomputing Network. We thank L. Valdés for technical assistance and Nick Guthrie for his excellent text editing.